\documentstyle[preprint,eqsecnum,aps]{revtex}
\def\beq{\begin{equation}} \def\eeq{\end{equation}}
\def\beqa{\begin{eqnarray}} \def\eeqa{\end{eqnarray}} 
\def\noeq{\nonumber} \def\disp#1{\displaystyle{#1}}
\def\del{\delta} \def\eps{\varepsilon}
\def\half{\frac{1}{2}}  
\def\mpi{m_\pi}  
\def\vp{\mbox{\boldmath$p$}} \def\vq{\mbox{\boldmath$q$}}
\def\vk{\mbox{\boldmath$k$}} \def\vS{\mbox{\boldmath$S$}}
\def\vt{\mbox{\boldmath$t$}} 
\def\veps{\mbox{\boldmath{$\varepsilon$}}}
\def\vs{\mbox{\boldmath{$s$}}} \def\vmu{\mbox{\boldmath{$\mu$}}}
\def\vJ{\mbox{\boldmath{$J$}}} \def\hvq{\hat{\vq}}
\def\hvs{\hat{\vs}} \def\vsig{\mbox{\boldmath{$ \sigma$}}}
\def\omk{\omega_k} \def\omq{\omega_q}
\def\CT{{\cal T}}  \def\CP{{\cal{P}}}

\def\const{\left(\frac{ef_{Y'Y\pi}}{4\pi}\right)}
\def\GLP{\left(\frac{ef_{\Lambda\Sigma\pi}}
{8\pi\mpi M_N}\right)\mu^V_{\Sigma\Lambda}}

\def\GXL{\left(\frac{ef_{\Sigma^+\Xi^0 K^+}}{8\pi m_KM_N}\right)
\mu^V_{\Sigma\Lambda}}

\def\DM{\Delta_{\Sigma\Lambda}} 
\def\DML{\Delta_{\Sigma^*\Lambda}}
\def\DMS{\Delta_{\Sigma^*\Sigma}} 
\def\DMES{\Delta_{\Sigma^{**}\Sigma}}
\def\DMEL{\Delta_{\Sigma^{**}\Lambda}} 
\def\DESR{\Delta_{\Sigma^{**}\Sigma^*}}

 \def\XS{\Delta_{\Xi\Sigma}} 
 \def\XRS{\Delta_{\Xi^*\Sigma}} \def\XRX{\Delta_{\Sigma^*\Xi}}
\def\SRX{\Delta_{\Sigma^*\Xi}} \def\XES{\Delta_{\Xi^{**}_1\Sigma}} 
\def\XET{\Delta_{\Xi^{**}_3\Sigma}} 
\def\XEX{\Delta_{\Xi^{**}_1\Xi}} \def\XETX{\Delta_{\Xi^{**}_3\Xi}} 
\begin{document}
\draft
\preprint{
  \begin{tabular}{r}
    DPNU-99-34\\
    November 1999
  \end{tabular}
}

\title{\Large\bf One-loop calculations of hyperon polarizabilities \\
under the large $N_c$ consistency condition}

\author{Y. Tanushi\thanks{E-mail: tanushi@nuc-th.phys.nagoya-u.ac.jp}
    and S. Saito\thanks{E-mail: saito@nuc-th.phys.nagoya-u.ac.jp}}
\address{Department of Physics, Nagoya University,
Nagoya 464-8602, Japan}
\author{M. Uehara\thanks{E-mail: ueharam@cc.saga-u.ac.jp}}
\address{Department of Physics, Saga University,
Saga 840-8502, Japan}

 \date{\today}

\maketitle
\begin{abstract}
The spin-averaged electromagnetic polarizabilities of the 
hyperons $\Lambda$ and $\Sigma$ are calculated within the 
one-loop approximation by use of the dispersion theory. The 
photon and meson couplings to hyperons are determined so as 
to satisfy the large $N_c$ consistency condition. It is shown 
that in order for the large $N_c$ consistency condition to hold 
exotic hyperon states such as $\Sigma^{**}$  with $I=2$ and 
$J=3/2$ are required in the calculation of the magnetic 
polarizability of the $\Sigma$ state. 
\end{abstract}
\baselineskip 18pt
\section{Introduction}
Beyond the spin averaged  electromagnetic polarizabilities of 
the nucleon, the spin polarizabilities\cite{Ragusa} have recently attracted  
theoretical attention, because these quantities serve as a crucial test of 
the low energy effective theories.  Using the heavy baryon chiral 
perturbation theories (HBChPT)\cite{Bernard92a,Hemmert97,Hemmert98}, 
the spin polarizabilities have been calculated and compared with the 
multipole analyses\cite{Sandorfi94,HDT,Babusci98b}. The spin polarizabilities 
are also calculated by using the dispersion theory,  where the imaginary parts 
are given by the Born terms of the one pion photoproduction 
amplitude\cite{Lvov98,Tanushi99}.  The dispersion theory  with the Born terms 
is a method to calculate loop diagrams\cite{Lvov93,Saito95,Tanushi}, and it 
reproduces almost the same results by  HBChPT up to O($p^3$) or O($\eps^3$), 
but it includes partially higher chiral  order diagrams than O($p^3$);   for 
example  Ref. \cite{Tanushi99} gives  the forward spin polarizability 
$\gamma_0=-0.4\times 10^{-4}\,{\rm fm}^4$, while HBChPT up to 
O$(\epsilon^3)$  does 
$\gamma_0=2.0\times 10^{-4}\,{\rm fm}^4$\cite{Hemmert98}.

As to hyperons the spin-averaged polarizabilities  have so far been 
studied  in the quark model\cite{Lipkin92}, the SU(3) extension of the 
HBChPT results\cite{Bernard92c} and  the bound-state soliton 
model\cite{Gobbi96}, but the study of hyperon polarizabilities is quite 
insufficient, because the hyperon polarizabilities involve much more 
physical contents than that of the nucleon. Further, since measurements of 
$\gamma\Sigma$ interactions are planned\cite{Moinester}, 
detailed and comprehensive studies will be required.

In this paper we calculate the spin-averaged polarizabilities of the 
$\Lambda$ and $\Sigma$ hyperons within the one-loop approximation by 
applying the dispersion theory to the Compton scattering amplitude, where 
the imaginary parts of the amplitudes are given by the Born terms of the pion 
and kaon photoproduction amplitudes. The coupling constants of the 
photon and meson to the nucleon and hyperons in the meson photoproduction 
amplitudes are given by the spin-flavor symmetry which leads to the large 
$N_c$ consistency condition. We refer sometimes to the bound 
state approach to strangeness in the chiral soliton  
model\cite{Callan85,BoundKaon}, because it is an explicit model realizing  
the spin-flavor symmetry for baryons in large $N_c$ 
QCD\cite{Dashen94,Jenkins98},  and the results are shared with any large 
$N_c$ baryon theories  at leading  order in the $1/N_c$ expansion. We call 
the model as the bound kaon-soliton model(BKSM) hereafter.

The Born terms in the pion and kaon photoproduction amplitudes with the 
electric coupling of the photon, which we call the electric Born terms,
contribute 
to the electric and magnetic polarizabilities.  The polarizabilities by the
pion 
electric Born terms  are of O($N_c$) in the $1/N_c$ expansion,  while those by 
the  kaon electric Born terms  are of O($N_c^0$). 

The Born terms through the magnetic coupling of the photon, which we call the 
magnetic Born terms, also contribute to the magnetic polarizabilities.  The 
magnetic Born terms can interfere with the electric Born ones through the 
unitarity relation and contribute also to the magnetic polarizabilities.  The 
magnetic Born term is written as the sum of the spin 1/2  and 3/2 baryon poles,
each of which is of O($N_c^{3/2}$).  The large $N_c$ consistency condition 
leads to the cancellation among the  pole terms: The whole amplitude  reduces 
to O($N_c^{1/2}$), and as a result the  amplitude is finite at infinite
energies. 
We show that in order for the large $N_c$ consistency condition to work in the 
pion production process off the $\Sigma$ target the exotic  strange baryon 
state  denoted as $\Sigma^{**}$ with the isospin 2 and spin 3/2,  has to 
contribute to the amplitudes.  Similarly,  the condition requires two exotic 
states , $\Xi^{**}_1$ and $\Xi^{**}_3$ with isospin 3/2 and spin 1/2 and 3/2, 
respectively, for the kaon production amplitudes.  The necessity of such exotic 
states is common to the large $N_c$ baryon theories in order for the unitarity 
relations not to violate a definite $N_c$ dependence  of amplitudes, that is of 
O($N_c^{1-n/2}$) for $n-$meson reaction amplitudes\cite{Dashen94,Jenkins98}. 
The contributions from the magnetic Born terms are of the same order 
as those from the electric ones in the $1/N_c$ expansion, but the magnetic 
Born contributions partially go beyond the calculation of O($p^3$) chiral order 
diagrams in HBChPT.

This paper is organized as follows: We discuss the pion and kaon electric 
Born contributions in the next section.  The pion and kaon magnetic Born 
terms are given in Sec. III, and it is also discussed 
how the large $N_c$ consistency condition works with the exotic states.  
The conclusions and discussion are given in the last section.  

\section{Contributions from the pion and kaon electric Born terms} 
As stated in Introduction we use the dispersion integrals to compute the 
one-loop diagrams in the Compton scattering amplitudes, the imaginary 
parts of which are given through the electric Born terms.

\subsection{Pion loop contributions}
We start with the pion photoproduction amplitude, 
$T_a=\eps_\mu T^\mu_a$ for $\gamma+ Y\to\pi_a+Y'$ with $Y(Y')$ 
being the initial(final) hyperon with the strangeness $S=-1$,  
which is decomposed as 
\beq
T_a=t^{(-)}_aT^{(-)}+t^{(+)}_aT^{(+)}+t^{(0)}_aT^{(0)}+t^{(\del)}_aT^{(\del)},
\eeq
where each amplitude is a function of the pion momentum 
$\vq$ and photon  $\vk$, and the isospin factors are as follows: 
For $I=1$ channel such as $\gamma\Sigma\to\pi\Sigma$,
$t^{(-)}_a=i\eps_{a3b}I_b$, $t^{(+)}_a=\{I_a,I_3\}$, $t^{(0)}_a=I_a$, 
$t^{(\del)}_a=\del_{a3}I$
with $I_a$ being the conventional isospin matrix,  and $I$ the $3\times 3$ 
unit matrix, and for $\gamma\Lambda\to\pi\Sigma$, 
$t^{(-)}_a=i\eps_{a3b}\CT^{\Sigma\Lambda}_b$ and 
$t^{(+)}_a=0$ with $(\CT^{\Sigma\Lambda}_b)_{m0}=\del_{mb}$.

The electric Born term  of O($N_c^{1/2}$), that is of leading order in 
the $1/N_c$ expansion, is written model-independently as
\beq
T_E^{(-)}=\left(\frac{ef_{Y'Y\pi}}{4\pi\mpi}\right)
\left[i\vsig\cdot\veps+2i\vsig\cdot\vt
\frac{\veps\cdot\vq}{\mpi^2-(q-k)^2}\right],\label{eq:Eterm}
\eeq
where $\vt=\vk-\vq$, and $\veps$ is the polarization vector.  
Since other  electric Born terms $T^{(+,0)}_E$ 
are of O($N_c^{-1/2}$) and break the unitarity limit at high 
energy, we ignore them hereafter\cite{Saito95}.

The pseudovector pion coupling constants to the hyperons $f_{YY'\pi}$ are 
given as  
\beq
\frac{1}{\mpi}\frac{f_{Y'Y\pi}}{\sqrt{4\pi}}=\Lambda_{Y'Y}G_\pi
 \label{eq:fyy}
\eeq
in the large $N_c$ baryon model and in the BKSM\cite{Kondo96}, where the 
overall constant $G_\pi$ is given empirically 
in the former and given in terms of the chiral angle $F(r)$ 
of the Skyrme model in the latter. 
The  factor $\Lambda_{Y'Y}$  satisfies the following spin-flavor symmetry 
relation\footnote{The sign of 
$\Lambda_{\Sigma\Sigma}$ and $\Lambda_{\Sigma^*\Sigma}$ is different 
from those in \cite{Kondo96}, because  the sign of the isospin matrix 
for $I=1$ is changed to the usual one, here.}:
\beq
\Lambda_{\Lambda\Sigma}=-\Lambda_{\Sigma\Sigma}
=-\frac{1}{\sqrt{3}}\Lambda_{\Sigma^*\Lambda}=
-\frac{2}{\sqrt{3}}\Lambda_{\Sigma^*\Sigma}
=-\Lambda_{NN}=\frac{1}{3}. \label{eq:Lambda}
\eeq  
This relation will play an important role for the large $N_c$ consistency 
condition in the magnetic Born terms. 
To fix the pseudovector coupling constant we adopt 
$|f_{\Sigma\Lambda\pi}|/\sqrt{4\pi}=0.22$\cite{Kondo96}, which is 
close to the empirical value $0.20\pm 0.01$\cite{Dumbrajs}, 
and other coupling constants are obtained according to the $\Lambda$ 
factor in the above.

According to Ref.\cite{Lvov93,Saito95,Tanushi} the forward dispersion 
relation with use of the electric Born term is known to give the 
electromagnetic polarizabilities  as follows:
\beq
\left(\begin{array}{c}
\alpha_Y(Y')\\[0.2cm]
\beta_Y^E(Y') \end{array}\right)
=A_{Y'}\const^2\frac{1}{24\mpi^3}\left(\begin{array}{c}
f(d)\\[0.2cm] g(d)\end{array}\right), \label{eq:alphabeta}
\eeq
where the factor $A_{Y'}$ is the multiplicity coming  
from the sum over $a$ and the spin components. The functions $f(d)$ and 
$g(d)$  are defined as for $d>-1$ and $d\ne 1$ with $d=(M_{Y'}-M_Y)/\mpi$
\beq
\left(\begin{array}{c}f(d)\\[0.2cm]g(d)\end{array}\right)=
\frac{2}{\pi}
\left(\begin{array}{c}
\displaystyle{d-2A(d)+\frac{9(d+2A(d))}{(d^2-1)}} \\[0.3cm] 
-(d+2A(d)) \end{array}
\right) , \label{eq:fdgd}
\eeq 
where
\begin{equation}
A(d)=\left\{
 \begin{array}{ll}
  \displaystyle{{1\over\sqrt{1-d^2}}\left[\tan^{-1}{d\over\sqrt{1-d^2}}
   -\tan^{-1}{1+d\over\sqrt{1-d^2}}\right]} &
  (|d|<1) \\[0.3cm]
  \displaystyle{-{1\over 2\sqrt{d^2-1}}\log\left(d+\sqrt{d^2-1}\right)} &
  (d>1)
 \end{array}
\right . ,
\end{equation}
and for $d=1$ we have $ f(d)=16/\pi,\, g(d)=0$.
In the above $\Sigma^*(1385)$ is also included in $Y'$.  
We use  the  empirical mass spectrum for the  
hyperons, the nucleon and $\Delta$ throughout  the paper. 
The calculated results of the electromagnetic polarizabilities from the pion
electric Born terms are tabulated in Table \ref{tb:table1}.

We observe from Table \ref{tb:table1} that the electric polarizability of 
the hyperons are in order 
 \beq
 \alpha_{\Sigma^{\pm}}\,>\,\alpha_\Lambda\,>\,\alpha_{\Sigma^0}. 
 \eeq
Due to the large coupling constant $f_{\Sigma^*\Lambda\pi}$, the 
contribution from $\Sigma^*$ to the $\Lambda$ electric polarizability is rather 
large, similar to the nucleon case.  We note that the effect by the mass 
difference among the hyperons is rather significant as seen in the difference 
between the $\Lambda\pi$ contribution in the $\Sigma$ target and $\Sigma\pi$ 
in the $\Lambda$ target, where the former is exothermal and the latter 
endothermal. The difference between 
$\alpha_{\Sigma^+}$ and $\alpha_{\Sigma^-}$ cannot be calculated within this 
approximation.

\subsection{Kaon loop contributions}
We obtain the kaon electric Born term of  O($N_c^0$) at leading order  
in the $1/N_c$ expansion  for $\gamma+Y\to \bar K_\alpha(K_\alpha)+B$ 
with $B$ being $N$ and $\Delta$ ($\Xi$ and $\Xi^*$) by replacing  $\mpi$  
by $m_K$ and the coupling constant $f_{Y'Y\pi}$ by $f_{YBK}$
in Eq.(\ref{eq:Eterm}).  

The P-wave kaon coupling constant $f_{YBK}$ is of O($N_c^0$) and given as
\beq
\frac{1}{m_K}\frac{f_{YBK}}{\sqrt{4\pi}}=\Lambda_{YBK}G_K,
\label{eq:kaoncc}
\eeq
and we fix  the pseudovector kaon coupling constant of 
$\Lambda pK^-$ as $f_{\Lambda pK}/\sqrt{4\pi}= 0.92$\cite{Kondo96},
while the empirical one is $0.89\pm 0.10$.  The value 0.92 in the kaon mass 
scale as in Eq.(\ref{eq:kaoncc}) is  reduced to 0.26 in the case of the pion 
mass scale, that is of the same order as the pion coupling constant.  The 
large $N_c$ relation of $\Lambda_{YBK}$ for the charged kaons is given as 
\beq
\Lambda_{\Lambda pK^-}=\sqrt{3}\Lambda_{\Lambda\Xi^-K^+}
=-\frac{1}{2}\Lambda_{\Lambda\Xi^{*-}K^+}=-\frac{1}{\sqrt{2}}
\eeq
for $\Lambda$ vertices, and  
\beq
\Lambda_{\Sigma^- nK^-}
=-\frac{1}{\sqrt{3}}\Lambda_{\Sigma^-\Delta^0K^-}
=-\frac{\sqrt{3}}{5}\Lambda_{\Sigma^+\Xi^0K^+}
=-\frac{1}{2}\Lambda_{\Sigma^+\Xi^{*0}K^+}
=-\frac{1}{3}
\eeq
for $\Sigma$ vertices.

The kaon contributions are given as 
\beq
\left(\begin{array}{c}
\alpha_Y\\[0.3cm]
\beta_Y^E
\end{array}\right)=
\sum_B A_B\left(\frac{ef_{YBK}}{4\pi}\right)^2
\frac{1}{24 m_K^3}
\left(\begin{array}{c}
f(d)\\[0.3cm]
g(d)
\end{array}\right),
\eeq
where $A_B$ is the same multiplicity as $A_{Y'}$ in the pion production, and 
$d=(M_Y-M_B)/m_K$. Note that the factor $(m_\pi/m_K)=0.279$ reduces the 
size of the kaon contributions. The numerical results from the
kaon electric Born terms are tabulated in Table \ref{tb:table2}.

The kaon-loop contributions to  polarizabilities are in order
\beq
\alpha_\Lambda\,>\,\alpha_{\Sigma^+}\,>\, \alpha_{\Sigma^0}\,
>\,\alpha_{\Sigma^-}.
\eeq
We see that the contributions from the decuplet baryons in the final 
states are of the same order as  those from the octet baryons.

The kaon contribution leads to the result $\alpha_{\Sigma^+}\,
>\,\alpha_{\Sigma^-}$, 
that is of the same sign in Ref.\cite{Bernard92c}. At the same time the 
kaon contribution to the nucleon makes the proton polarizabilities larger 
than  the neutron ones, but it does not agree with
the experimental tendency: $\alpha_n\,>\,\alpha_p$.

\section{Magnetic Born terms and the large $N_c$ consistency 
condition}
The spatial part of the electromagnetic current, $\vJ$, 
contributes to the magnetic Born term, where   
\beq
<Y'(\vp')|\veps\cdot\vJ|Y(\vp)>=<Y'|i\vs\cdot\vmu|Y>
\eeq
with $\vs=(\vp'-\vp)\times\veps$ and $\vmu$ being the magnetic moment 
operator. The magnetic moment is decomposed as 
\beq
<Y'|\vmu|Y>=\vS\frac{e}{2M_N}(\mu^V_{Y'Y}\CT_3+\mu^S_{Y'Y}),
\eeq
where $\vS(\CT_3)$ is the transition spin (isospin) matrix, and 
$\mu^{V}_{Y'Y}\,(\mu^{S}_{Y'Y})$ is the isovector (isoscalar) part of the 
hyperon magnetic moment in units of  Bohr magneton. 
Since the isovector part, $\mu^V e/2M_N$, is of O($N_c$), while 
the isoscalar part is of  O($N_c^0$), the leading contributions come 
from the isovector part of the magnetic moments. We further note that 
$\mu^V_{Y'Y}$ having the same strangeness $S$ is proportional to the 
factor $\Lambda_{Y'Y}$ of Eq.(\ref{eq:Lambda})\cite{Dashen94,Kunz,Oh}.

The magnetic Born term for a process $\gamma+Y\to \pi^a+Y'$ is written 
as
\beq
T_M^a=\sum_{Y''}\left\{\left(\frac{ef_{Y'Y''\pi}}{8\pi\mpi M_N}\right)
\mu^V_{Y''Y}
\frac{(\vS'\cdot\vq)^\dagger(\vS\cdot\vs){\CT'_a}^\dagger\CT_3}
{M_{Y''}-M_{Y'}-\omq}+\left(\frac{ef_{Y''Y\pi}}{8\pi\mpi M_N}\right)
\mu^V_{Y'Y''}
\frac{(\vS'\cdot\vs)^\dagger(\vS\cdot\vq){\CT'_3}^\dagger\CT_a}
{M_{Y''}-M_Y+\omq}\right\}
\eeq
at leading order of the O($1/N_c$) expansion, where $\vS'(\CT_a')$ is  
the transition spin(isospin) matrix for the $Y''\to Y'$ vertex, 
while those without a prime for the $Y\to Y''$ vertex. 
The magnetic Born terms are decomposed as 
\beq
T^a_M=\sum_{\ell=1,3}\sum_{n=\pm,\del}\CP_\ell(\hvq,\hvs) \, 
t^{(n)}_a\,T^{(n)}_\ell(\omq), 
\eeq
where   $\CP_1(\CP_3)$ is the projection operator for the P-wave pion 
production amplitude with  total angular momentum $j=1/2(3/2)$:
\beq
\CP_1(\hvq,\hvs)=(\vsig\cdot\hvq)(\vsig\cdot\hvs),\quad\mbox{and}\quad
\CP_3(\hvq,\hvs)=3(\hvq\cdot\hvs)-(\vsig\cdot\hvq)(\vsig\cdot\hvs)
\eeq
with  $\hat{\vq}=\vq/q$ and $\hat{\vs}=\vs/k$ for $Y'$ with spin 1/2. 
Similar expressions are written for  $Y'$ with spin 3/2.  The final states are 
restricted to the states with an octet or decuplet baryon accompanied with  
a pion or kaon.

We notice here that  in order for the large $N_c$ consistency 
condition to hold the exotic hyperon states are required for the $\Sigma$ 
target. Due to the consistency condition the magnetic Born terms  keep being 
of O($N_c^{1/2}$)  and are convergent at infinite energies, as a result. 
So, we concentrate ourselves to the $\Sigma$ target both for the pion and 
kaon magnetic Born terms in the following.

\subsection{The pion magnetic Born terms}
Here we discuss explicitly  the magnetic Born terms of the process 
$\gamma+\Sigma\to\pi^a+\Sigma$. 
Using the mass abbreviation $\Delta_{Y'Y}=M_{Y'}-M_Y$, we write 
the non-exotic pole amplitudes with the $\Sigma$ and $\Sigma^*$ poles 
in this channel as 
\beqa
T^{(\pm)}_1&=&\left(\frac{ef_{\Sigma\Sigma\pi}}{8\pi\mpi M_N}\right)
\mu^V_{\Sigma}\left[-\frac{kq}{2\omq}\mp\frac{kq}{6\omq}\right]
\pm\frac{1}{3}\left(\frac{ef_{\Sigma^*\Sigma\pi}}{8\pi\mpi M_N}
\right)
\mu^V_{\Sigma^*\Sigma}\left[\frac{2}{3}\frac{kq}{\DMS+\omq}\right],\\
T^{(\pm)}_3&=&\left(\frac{ef_{\Sigma\Sigma\pi}}{8\pi\mpi M_N}\right)
\mu^V_{\Sigma}\left[\pm\frac{kq}{3\omq}\right]
+\frac{1}{3}\left(\frac{ef_{\Sigma^*\Sigma\pi}}{8\pi\mpi M_N}\right)
\mu^V_{\Sigma^*\Sigma}\left[\frac{1}{2}\frac{kq}{\DMS-\omq}\pm
\frac{1}{6}\frac{kq}{\DMS+\omq}\right],
\eeqa
and those with  the $\Lambda$ pole are similarly written, but not 
shown explicitly. We note that each pole term is of O($N_c^{3/2}$).

If we use the relation given by the $\Lambda$ factors,  
\beq
f_{\Sigma\Sigma\pi}\mu^V_{\Sigma}=
\frac{4}{3}f_{\Sigma^*\Sigma\pi}\mu^V_{\Sigma^*\Sigma}
=f_{\Sigma\Lambda\pi}\mu^V_{\Sigma\Lambda},
\eeq
we see that the cancellation  does not occur for  the sums of the above 
amplitudes. Notice, however, that the $\Sigma+\pi$ channel can 
communicate with the $I=2$ channel 
with strangeness $-1$.   The large $N_c$ baryon theories and BKSM  
predict two such exotic baryons  with spin 3/2 and 5/2, where the 
P-wave antikaon is bound around the soliton with isospin 2 in the 
latter model. In this channel we need the exotic state with 
$I=2$ and spin 3/2 in order for the large $N_c$ consistency 
condition to hold.  (The exotic state with spin 5/2 cannot interact 
with P-wave $\pi\Sigma$ states.)  
We denote the  exotic state with spin 3/2 as $\Sigma^{**}$.
Including the exotic state, 
we can see that the cancellation occurs with the spin-flavor symmetry
relation 
\beq
f_{\Sigma^{**}\Sigma\pi}\mu^V_{\Sigma^{**}\Sigma}=
\frac{3}{2}f_{\Sigma^*\Lambda\pi}\mu^V_{\Sigma^*\Lambda}=
\frac{9}{2}f_{\Sigma\Sigma\pi}\mu^V_{\Sigma}=
\frac{9}{2}f_{\Sigma\Lambda\pi}\mu^V_{\Sigma\Lambda}.
\eeq
Indeed, BKSM gives  the factor 
$\Lambda_{\Sigma^{**}\Sigma}$ as 
\beq
\Lambda_{\Sigma^{**}\Sigma}=-\frac{1}{\sqrt{2}},
\eeq
that is consistent with the above condition for the cancellation, of 
course. This result is shared with the large $N_c$ baryon theories.

We summarize the resultant Born terms as 
\beqa
T^{(-)}_1&=&\GLP\cdot kq
\left[-\frac{1}{6}\frac{\DML}{(\DMS+\omq)(\DM-\omq)}
-\frac{1}{3}\frac{\DMES}{\omq(\DMES+\omq)}\right.\noeq\\
&-&\left.\frac{1}{2}\frac{(\DMES-\DM)}{(\DM+\omq)(\DMES+\omq)}\right],\\
T^{(+)}_1&=&\GLP\cdot kq
\left[-\frac{1}{2}\frac{\DM}{\omq(\DM+\omq)}
-\frac{1}{6}\frac{\DMS}{\omq(\DMS+\omq)}\right.\noeq\\
&+&\left.\frac{1}{6}\frac{\DMEL}{(-\DM+\omq)(\DMES+\omq)}\right],\\
T^{(\del)}_1&=&\GLP\cdot kq
\left[-\frac{\DMES-\DM}{(\DM+\omq)(\DMES+\omq)}+
\frac{1}{3}\frac{\DMEL}{(\DM-\omq)(\DMES+\omq)}\right],\\
T^{(-)}_3&=&\GLP\cdot kq
\left[\frac{1}{4}\frac{\DMS}{({\DMS}^2-\omq^2-i\DMS\Gamma_{\Sigma^*})}-
\frac{5}{4}\frac{\DMES}{({\DMES}^2-\omq^2-i\DMES\Gamma_{\Sigma^{**}})}
\right.\noeq\\
&-&\left.\frac{1}{3}\frac{\DMES}{\omq(\DMES+\omq)}
-\frac{1}{6}\frac{\DESR}{(\DMS+\omq)(\DMES+\omq)}
-\frac{1}{3}\frac{\DMEL}{(-\DM+\omq)(\DMES+\omq)}\right],\\
T^{(+)}_3&=&\GLP\cdot kq
\left[\frac{1}{4}\frac{\DMS}{({\DMS}^2-\omq^2-i\DMS\Gamma_{\Sigma^*})}-
\frac{1}{4}\frac{\DMES}{({\DMES}^2-\omq^2-i\DMES\Gamma_{\Sigma^{**}})}
\right.\noeq\\
&+&\left.\frac{1}{3}\frac{\DM}{\omq(\DM-\omq)}
-\frac{1}{12}\frac{\DESR}{(\DMS+\omq)(\DMES+\omq)}\right],\\
T^{(\del)}_3&=&\GLP\cdot kq 
\left[\frac{2\DMES}{{\DMES}^2-\omq^2
-i\DMES\cdot\Gamma_{\Sigma^{**}}}
-\frac{2}{3}\frac{\DMEL}{(\DM-\omq)(\DMES+\omq)}\right].
\eeqa
Since the mass differences of O($N_c^{-1}$) appear in the numerators, the 
resultant amplitudes are  reduced to  O($N_c^{1/2}$) and finite at high 
energies. 
We stress here that we have not introduced any vertex functions depending 
on the meson or photon momentum, because the vertex corrections go beyond 
the one-loop approximation. In this sense the exotic states play a role similar 
to a natural cutoff function without destroying the analytic structure of 
the one-loop amplitudes. 

In the above we  inserted the total width $\Gamma_{\Sigma^*}$ and 
$\Gamma_{\Sigma^{**}}$ into the pure resonance terms of $T_3$. 
The width $\Gamma_{\Sigma^*}$ is given as 
\beq
\Gamma_{\Sigma^*}=\frac{2}{3}\left(
\frac{f_{\Sigma^*\Lambda\pi}^2}{4\pi}\right)
\frac{q_\Lambda^3}{\mpi^2}+\frac{4}{3}\left(
\frac{f_{\Sigma^*\Sigma\pi}^2}{4\pi}\right)
\frac{q_{\Sigma}^3}{\mpi^2}, 
\eeq
where $q_{\Lambda}(q_{\Sigma})$ is the pion momentum decaying into 
the channel $\Lambda(\Sigma)+\pi$. This form of the width is the same 
as that adopted in the previous works\cite{Tanushi99,Tanushi}, 
which guarantees 
the narrow width limit. Numerical values for the 
widths including recoil are $\Gamma_{\Sigma^*}^{\Lambda\pi}=44.8$ MeV 
and $\Gamma_{\Sigma^*}^{\Sigma\pi}=4.7$ MeV.  Similar form of the 
width of $\Sigma^{**}$ is used, where $\Sigma^{**}$ is supposed to 
decay only to $\Sigma\pi$ channel, because the $\Sigma^{**}$ 
mass is not expected to be so higher than the $\Sigma^*\pi$ threshold as 
seen below.  Then, we have $\Gamma_{\Sigma^{**}}^{\Sigma\pi}=138$ MeV,  
but since $\Lambda+2\pi$ channel is not taken into account, though it opens,  
the width of  $\Sigma^{**}$ would be underestimated.

Since all the isovector magnetic moments appearing in the 
amplitudes are for the hyperons with $S=-1$, we take the 
empirical value, $\mu^V_{\Sigma\Lambda}=-1.61$, 
to fix the magnetic coupling constants.  
The isoscalar magnetic moment  can  give  pole terms 
of O($N_c^{1/2}$), but they are not of leading order. Since the cancellation 
among the pole terms does not hold at non-leading order and then the 
unitarity bound is broken, we disregard them as in the case of the 
electric pion Born terms.

\subsection{The kaon magnetic Born terms} 
Contrary to the pion photoproduction processes,  the isovector 
magnetic moments with different strangeness from $S=0$ to $-2$, 
contribute to the $\bar K$ or $K$ photoproduction amplitudes. 
Although the magnetic moment is not completely proportional to the 
$\Lambda$ factor,
we take the experimental value for $\mu^V_{\Sigma\Lambda}$,
by which the other magnetic moments are fixed,
because it may be regarded as giving an average.

In the case of the   $\gamma +\Sigma\to K+\Xi$ process,  the 
cancellation does not  occur among the $\Xi$ and 
$\Xi^*$ pole terms: The two exotic $\Xi$ states with isospin 3/2 contribute 
to this processes,  the one with  spin 1/2 denoted as $\Xi^{**}_1$  and the 
other with spin 3/2 as  $\Xi^{**}_3$.  The resultant amplitudes are written 
as 
\beqa
T_1&=&\GXL\cdot kq\left[\frac{16}{45}
\frac{\XRX}{(\XS+\omq)(\XRS+\omq)}
\right.\noeq\\
&+&\left.\frac{4}{45}\frac{\XEX}{(\XS+\omq)(\XES+\omq)}
+\frac{4}{9}\frac{\XETX}{(\XS+\omq)(\XET+\omq)}\right], \\
T_3&=&\GXL\cdot kq
\left[\frac{1}{5}\frac{\DMS}{(\XS+\omq)(\SRX-\omq)}
\right. \noeq\\
&+&\left.\frac{4}{45}\frac{\XRX}{(\XS+\omq)(\XRS+\omq)}
-\frac{8}{45}\frac{\XEX}{(\XS+\omq)(\XES+\omq)}
+\frac{1}{9}\frac{\XETX}{(\XS+\omq)(\XET+\omq)}\right]
\eeqa
for $\gamma+\Sigma^+\to K^++\Xi^0$, and similar amplitudes are written 
for other $\Sigma$ targets. In the above  we used 
\beq
\Lambda_{\Sigma^+\Xi^{**0}_1K^+}=\frac{\sqrt{2}}{3\sqrt{3}} 
\quad{\rm and}\quad  
\Lambda_{\Sigma^+\Xi^{**0}_3K^+}=\frac{\sqrt{5}}{3}, \label{eq:exoticXi}
\eeq
which are predicted by BKSM. Each pole term in 
the above is of O$(N_c)$, but the resultant amplitudes are of O$(N_c^0)$, 
because the mass differences appearing in the numerators are of O$(N_c^{-1})$. 
The large $N_c$ consistency condition also works to reduce the $N_c$ 
dependence and to converge the asymptotic behavior. Since the resonance 
poles do not develop in the physical region of $\omq$, their contributions are 
rather small.

\subsection{Mass spectrum of the exotic states}
The exotic states are required to satisfy the large $N_c$ consistency 
condition of the production amplitudes as shown in  previous subsections. 
Here we estimate the masses within BKSM. The baryon mass spectrum of the 
model is given by the following formula as 
\beqa
M=M_s+|s|\omega&+&\frac{1}{2\Lambda}\left[cJ(J+1)+(1-c)I(I+1)
-c(1-c) \frac{|s|}{2}\left(\frac{|s|}{2}+1\right)\right],
\eeqa
where $M_s,\omega,c$ and $\Lambda$ are the parameters to be calculated 
by the model\cite{BoundKaon,Kunz}. Instead of calculating these 
parameters by the model we estimate them by the existing mass  spectrum 
of the non-strange and strange baryons: The result we adopt is  
$M_s=866$ MeV, $c=0.630$ , $\omega=221$ MeV and 
$M_\Delta-M_N=3/(2\Lambda)=293$ MeV.
The same parameters give the masses  of the exotic states as 
\beq
\begin{array}{rcl}
\disp{\Sigma^{**}(I=2,J=\frac{3}{2})}&=&1517\,{\rm MeV},\\[0.2cm]
\disp{\Xi^{**}_1(I=\frac{3}{2},J=\frac{1}{2})}&=&1444\,{\rm MeV},
\qquad
\disp{\Xi^{**}_3(I=\frac{3}{2},J=\frac{3}{2})}= 1639\,{\rm MeV}.
\end{array}
\eeq
We note that almost the 
same values are obtained for the masses of the exotic states by the mass 
formula in the tree level of the large $N_c$ chiral perturbation 
theory\cite{OhWeise}.

The mass of $\Sigma^{**}$ is high enough  to decay into $\Sigma\pi, 
\Sigma\pi\pi$ and $\Lambda\pi\pi$ channels, while the mass of 
$\Xi^{**}_1$ is low and seems to be stable. It should 
be noticed, however, that the exotic states could disappear at $N_c=3$. 
The masses of the exotic states may be sensitive to higher $N_c$ 
corrections, but we use 1520 MeV for $\Sigma^{**}$, 1450 (1640) MeV 
for $\Xi^{**}_1(\Xi^{**}_3)$  in this paper.

\subsection{Magnetic polarizability from the magnetic Born terms}
The magnetic polarizability $\beta^M_Y$  is given 
by the integration  over energy as follows:
\beq
\beta^M_Y=\frac{2}{\pi}\int^\infty_{\omega_{\rm th}}
\frac{d\omk}{\omk^2}\frac{q}{\omk}
 \sum_{m,n}\left({T_1^{(n)}}^*T_1^{(m)}+2{T_3^{(n)}}^*T_3^{(m)}\right)
 \sum_a{t_a^{(n)}}^\dagger t_a^{(m)}
\label{eq:betaM}
\eeq
for the spin 1/2 final baryon. 

The magnetic Born term can interfere with the electric one: By the angular 
integration  we have 
\beq
\beta^I_Y=\sum_{Y'}\frac{2G_E}{\pi}\int \frac{d\omk}{\omk^2}
\frac{q}{\omk}
\left[\frac{1}{v}-\frac{1-v^2}{2v^2}\log\left(\frac{1+v}{1-v}\right)\right]
\sum_{n,a}\left({\rm Re}T^{(n)}_1-{\rm Re}T^{(n)}_3\right)
{t_a^{(-)}}^\dagger t_a^{(n)} \label{eq:betaI}
\eeq
for the spin 1/2 final baryon, where $G_E$ denotes the corresponding coupling 
constant in $T_E$.  In the case of the spin 3/2 final baryon, the above
expressions are little changed. 
In Table\,\ref{tb:table3} we show the numerical results of the magnetic
polarizabilities, in which all the contributions are included.

Instead of integrating the full amplitudes, if we pick up only the the 
$\Sigma^*$ state and ignore the exotic state and the background 
contributions at all, we may get rid of the contributions from the exotic 
state. Such a narrow width approximation has been 
discussed in the case of the nucleon polarizabilities and shown to give 
the same result as the one by  HBChPT in Ref.\cite{Tanushi99,Tanushi}. 
So, we proceed to the narrow width approximation for the 
$\gamma+\Lambda\to \Lambda+\pi$ channel as a typical example: 
$|T^{(\pm)}_3|^2$ contains the $\Sigma^*$ resonance and its contribution 
in Eq.(\ref{eq:betaM}) is proportional to  
\beq
\frac{e^2}{4\pi}\left(\frac{\mu^V_{\Sigma\Lambda}}{2M_N}\right)^2
\frac{2}{\pi}\int\frac{d\omk}{\omk^3}
\left(\frac{f_{\Sigma\Lambda\pi}^2}{4\pi}\frac{q^3}{\mpi^2}\right)
\frac{4{\DML}^2}{({\DML}^2-\omq^2)^2+(\DML\,\Gamma_{\rm tot})^2},
\eeq
where $\Gamma_{\rm tot}$ is the total width of $\Sigma^*$. 
Using  the relation 
\beq
\frac{f_{\Sigma\Lambda\pi}^2}{4\pi}\frac{q^3}{\mpi^2}=\half\left[
\frac{2}{3}\frac{f_{\Sigma^*\Lambda\pi}^2}{4\pi}\frac{q^3}{\mpi^2}
\right]=\half\Gamma_{\Lambda\pi},
\eeq
and  taking  the narrow width limit as 
\beq
\lim_{\Gamma_{\rm tot}\to 0}
\frac{\DML\Gamma_{\rm tot}}{({\DML}^2-\omq^2)^2+
(\DML\,\Gamma_{\rm tot})^2}=\pi\del({\DML}^2-\omq^2), 
\eeq
we may have 
\beq
\beta^M_{\Lambda\to\Lambda}|_{\Sigma^*}=\frac{e^2}{4\pi}\left(
\frac{\mu^V_{\Sigma\Lambda}}{2M_N}\right)^2
\frac{4}{\DML}\left(
\frac{\Gamma_{\Lambda\pi}}{\Gamma_{\rm tot}}\right),
\eeq
where the spin factor 2 is multiplied and the last factor is the 
branching ratio of $\Sigma^*\to \Lambda\pi$. Adding the 
$\Sigma^\pm\pi^\mp$ channels, we get 
\beq
\beta^M_{\Lambda}|_{\Sigma^*}=\left(\frac{e^2}{4\pi}\right)
\left(\frac{\mu^V_{\Sigma\Lambda}}{2M_N}\right)^2
\frac{4}{\DML},
\eeq
which is similar to the case of the nucleon.  For the $\Sigma$ target 
we have 
\beq
\beta^M_{\Sigma^\pm}|_{\Sigma^*}=
\left(\frac{e^2}{4\pi}\right)\left(
\frac{\mu^V_{\Sigma\Lambda}}{2M_N}\right)^2
\frac{1}{\DMS},\qquad
\beta^M_{\Sigma^0}|_{\Sigma^*}=0.
\eeq
Since $\Sigma^0$ does not have the leading isovector 
magnetic moment, $\beta_{\Sigma^0}|_{\rm narrow}$ is zero  
at leading order.  Even if the isoscalar magnetic moment is used, it is at
most 1/4 of the $\beta_{\Sigma^\pm}$, because 
$\mu^S\approx 1/2\mu^V_\Sigma$. The numerical results are as follows: 
$$ \beta^M_\Lambda|_{\Sigma^*}=6.13\quad{\rm and}\quad
\beta^M_{\Sigma^\pm}|_{\Sigma^*}=2.15$$
in units of $10^{-4}\,{\rm fm}^3$.

Similar narrow width approximation to $\Sigma^{**}$  gives the values, 
$$\beta^M_{\Sigma^\pm}|_{\Sigma^{**}}=3.78\quad{\rm and}\quad
\beta^M_{\Sigma^0}|_{\Sigma^{**}}=5.04.$$  Since  
$\Gamma_\Sigma^{**}$ is broad as seen  previously, these values 
would be an overestimate, but the exotic resonance contributions cannot be   
discarded, especially to $\Sigma^0$. 

Finally, we note that $\beta$ in the narrow width approximation  
is of O($N_c^3$), because the limit picks up only the relevant pole of 
O($N_c^{3/2}$); that is, the limit is not consistent with the $1/N_c$
expansion.

\section{Conclusions and discussion}
We have calculated the spin-averaged electromagnetic polarizabilities 
of the hyperons $\Lambda$ and $\Sigma$ within the one-loop approximation.   
In order to calculate the one-loop diagrams we used the dispersion relations, 
where the imaginary parts  are given by the Born terms in the pion and kaon 
photoproduction Born terms. The  Born terms satisfy the 
low energy theorems, and their form is  model-independent. The coupling 
constants are determined so as to satisfy spin-flavor symmetry of the large 
$N_c$ QCD.

The calculated electromagnetic polarizabilities through the pion and kaon Born 
terms are summarized as $\alpha_\Lambda=18.05$, $\alpha_{\Sigma^+}=22.08$, 
$\alpha_{\Sigma^0}=13.79$ and $\alpha_{\Sigma^-}=18.71$ , and 
$\beta_\Lambda=3.22$, $\beta_{\Sigma^+}=6.67$, $\beta_{\Sigma^0}=5.52$ 
and $\beta_{\Sigma^-}=7.13$ in units of $10^{-4}\,{\rm fm}^3$.  These values 
would be too large as seen from the large values of the polarizabilities of the 
nucleon given by the same calculation\cite{Tanushi} as well as  
the one-loop calculation in HBChPT\cite{Hemmert97}. 
This is because the high energy contributions from the one-loop diagrams 
are not fully reduced for the spin-averaged polarizabilities compared to
the spin 
polarizabilities owing to the power behavior  of the energy denominator in the 
integrals.  In order to reduce the values of the spin-averaged polarizabilities 
within the one-loop calculations we would have to go to the approximation 
beyond the one-loop level, such as vertex corrections and the unitarization
of the Born amplitudes. 

The electric Born terms would give the same spin-averaged 
polarizabilities as the SU(3) extension of HBChPT\cite{Bernard92c}, 
if the hyperon mass differences are ignored.  But we observed that the 
polarizabilities  strongly depend on the hyperon mass difference, and then 
SU(3) symmetry  of the polarizabilities is further broken  besides the symmetry 
breaking due to the pion and kaon mass difference, even if the coupling 
constants satisfy SU(3) symmetry with an appropriate $F/D$  ratio.

As to the magnetic Born amplitudes we have shown that exotic hyperon 
states are inevitably required even in the non-exotic reaction channel in order 
for the large $N_c$ consistency condition to hold. The consistency condition  
guarantees meson-baryon reaction amplitudes to have a meaningful $N_c$ 
limit. Due to the consistency condition the magnetic 
Born terms remain at O($N_c^{1/2}$),  and as a result they 
become finite at high energies and give finite magnetic polarizabilities 
as  the electric Born ones.   We also  noted that the narrow width 
limit is not consistent with the $1/N_c$ expansion, because the limit picks 
up the single resonance pole term of O($N_c^{3/2}$) and the resultant 
polarizability is of O($N_c^3$).  If we reduce $N_c$ to 3, the coupling 
constants of the exotic hyperons to non-exotic ones would 
vanish, but simultaneously it makes the magnetic Born amplitudes break the 
unitarity bound at high energies even for such a case of the nucleon.
 Thus, it is impossible to have finite results within  the one-loop
approximation. 
Contrary, it is a serious problem for the large $N_c$ 
baryon theories to study the $1/N_c$ corrections to physical quantities 
related to the exotic states and what physical effects  are expected by 
the exotic states besides a role of the natural cutoff,  if the leading
terms in 
the $1/N_c$ expansion are valid. These tasks are left to further
investigations. 

It is known that there is a negative parity hyperon $\Lambda^*(1405)$, 
which  BKSM  predicts as an S-wave bound state 
of the antikaon around the $I=J=0$ chiral SU(2) soliton.  The model also 
predicts $\Sigma_{1/2^-}$ and $\Sigma_{3/2^-}$, which are the bound 
states to the $I=J=1$ soliton\cite{Kondo96}. The electric dipole 
transition amplitudes with the poles at the negative parity hyperons 
of spin 1/2 give the electric polarizabilities. Since the transition 
electric dipole moment is of O($N_c^0$) and the S-wave pion coupling 
constant of O($N_c^{-1/2}$),  the electric polarizabilities are of 
O($N_c^{-1}$), compared to the electric Born contributions of O($N_c$). 
Taking the narrow width approximation given by  
$$\alpha^D=\frac{e^2}{4\pi}\left(
\frac{\kappa_{\Lambda^*Y}}{2M_N}\right)^2
\frac{2}{M_{\Lambda^*}-M_Y},$$ 
we get $\alpha^D_\Lambda=0.18$ and $\alpha^D_{\Sigma^0}=0.24$ in units 
of $10^{-4}\,{\rm fm}^3$,  where we used  the transition dipole moments 
$\kappa_{\Lambda^*\Lambda}=\kappa_{\Lambda^*\Sigma}= 0.41$
in units of the Bohr magneton, which are given 
by BKSM. The model also predicts that 
the dipole moment of $\Sigma^*_{1/2^-}$ is $-1/3$ of the $\Lambda^*$, 
and then the contribution to $\alpha$ would be tiny; $\alpha^D_{\Sigma^+}
=0.08$, $\alpha^D_{\Sigma^0}=0.02$ and $\alpha^D_{\Sigma^-}=0$.  
Our values are quite different from those of Ref.\cite{Lipkin92}, but 
of the same order as  Ref.\cite{Gobbi96}. 
The interference terms between the electric Born and the electric 
dipole moment terms are also small.  

Gobbi et al. calculated the polarizabilities of the hyperons in 
BKSM\cite{Gobbi96}, but they used the two-photon seagull 
terms in the Lagrangian.  It is, however, pointed out that it is dangerous 
to use the two-photon-seagull terms in the Lagrangian to calculate the 
polarizabilities, because the gauge invariance makes the seagull terms 
vanish for the electric polarizability\cite{Saito95,Lvov93Int,Saito98}.  
Although we referred to the same BKSM, our approach 
to the subject is quite different from theirs, and the results are also
different: 
We point out that the chiral soliton model including BKSM gives the 
model-independent form of the pion and kaon photoproduction amplitudes 
at tree level and then the polarizabilities are given by  calculating the 
loop integrals with the dispersion relations. 


\tightenlines
\begin{table}
\caption{
The electromagnetic polarizabilities from the pion electric  Born terms  
 in units of $10^{-4}\,{\rm fm}^3$.  $f_{\Sigma\Lambda\pi}/\sqrt{4\pi}$
 =0.22 is used to fix the pion coupling constants.
}
\begin{center}
\begin{tabular}{d|ddd|dddd|ddd}
$Y$ &&$\Lambda$ &&&& $\Sigma^{\pm}$ &
&& $\Sigma^0$& \\ 
$Y'$&$\Sigma$& $\Sigma^*$&total&$\Lambda$&$\Sigma$&$\Sigma^*$&total&
$\Sigma$&$\Sigma^*$&total\\ \hline
$\alpha_{Y}$&5.40&7.13&12.54&11.89&4.30&0.98&17.17&8.61&1.95&10.56\\ 
\hline
$\beta^E_{Y}$& 0.34&$-$1.29&$-$0.95&0.87&0.43&$-$0.07&1.23&
                    0.86&$-$0.14&0.72\\  
 \end{tabular}
 \end{center}
 \label{tb:table1}
 \end{table}
\begin{table}
\caption{The electromagnetic polarizabilities from the kaon electric Born 
terms in units of $10^{-4}\,{\rm fm}^3$. $f_{\Lambda pK^-}/\sqrt{4\pi}$
=0.92 is used to fix the kaon coupling constants.
}
\begin{center}
\begin{tabular}{d|ddd|ddd|ddd|ddd}
$Y$&&$\Lambda$&&&$\Sigma^+$&&&$\Sigma^0$&&&$\Sigma^-$&\\
$B$&$N+\Xi$& $\Xi^*$&total&$\Xi$&$\Delta+\Xi^*$&total&
 $N+\Xi$&$\Delta+\Xi^*$&total&$N$&$\Delta$&total\\ \hline
$\alpha_Y$& 3.16 &2.36&5.52& 2.36&2.56&4.91&1.62&1.61&3.23&0.88&0.67
&1.55\\ \hline
$\beta^E_Y$&0.28& 0.06&0.34&0.21&0.22&0.44&0.14&0.14&0.28&0.07
&0.07&0.13\\
 \end{tabular}
 \end{center}
 \label{tb:table2}
 \end{table}
\begin{table}
\caption{Total magnetic polarizability $\beta$ and each contribution 
$\beta^E_Y$, $\beta^M_Y$ or $\beta^I_Y$ in units of $10^{-4}\,{\rm fm}^3$. 
$\mu_{\Sigma\Lambda}^V$=$-$1.61 is used to fix the isovector part of 
hyperon magnetic moments. 
}
\begin{center}
\begin{tabular}{dd|d|d|d|d|d}
$Y$&total $\beta$&$\pi$ or $K$-loop & $\beta^E$&$\beta^M$& $\beta^I$ &
sum\\ \hline
$\Lambda$&3.22&$\pi+\Lambda,\Sigma$&0.34&5.36&$-$0.73&4.97\\
&&$\pi+\Sigma^*$ & $-$1.29&0.24&$-$1.25&$-$2.31\\
&&$\bar K+N,K+\Xi$ &0.28&1.04&$-$0.58&0.74\\
&&$K+\Xi^*$&0.06&0.07&$-$0.31&$-$0.18\\ \hline
$\Sigma^+$&6.67&$\pi+\Lambda,\Sigma$&1.30&5.08&0.24&6.61\\
&&$\pi+\Sigma^*$&$-$0.07&0.13&$-$0.20&$-$0.13\\
&&$\bar K+N,K+\Xi$&0.21&0.40&$-$0.34&0.27\\
&&$\bar K+\Delta,K+\Xi^*$&0.22&0.14&$-$0.45&$-$0.08\\ \hline
$\Sigma^0$&5.52&$\pi+\Sigma$& 0.86&4.06&0.46&5.38\\
&&$\pi+\Sigma^*$&$-$0.14&0.52&$-$0.42&$-$0.03\\
&&$\bar K+N,K+\Xi$&0.14&0.18&$-$0.19&0.13\\
&&$\bar K+\Delta,K+\Xi^*$&0.14&0.16&$-$0.25&0.05\\ \hline
$\Sigma^-$&7.13&$\pi+\Lambda,\Sigma$&1.30&5.08&0.24&6.61\\
&&$\pi+\Sigma^*$&$-$0.07&0.13&$-$0.20&$-$0.13\\
&&$\bar K+N,K+\Xi$&0.07&0.40&0.04&0.51\\
&&$\bar K+\Delta,K+\Xi^*$&0.07&0.14&$-$0.07&0.14\\
 \end{tabular}
 \end{center}
 \label{tb:table3}
\end{table}

\end{document}